\documentclass[12pt]{elsarticle}
\makeatletter
\def\ps@pprintTitle{%
	\let\@oddhead\@empty
	\let\@evenhead\@empty
	\def\@oddfoot{\centerline{\thepage}}%
	\let\@evenfoot\@oddfoot}
\makeatother
\usepackage{amsmath,amssymb,hyperref}
\usepackage{amssymb,amsmath,booktabs,hyperref}
\usepackage{graphics}
\usepackage{graphicx}
\usepackage{subfig}
\usepackage{multirow}
\usepackage{tabularx}
\usepackage{rotating}
\usepackage{lscape}
\usepackage{anysize}
\usepackage{xcolor}
\usepackage{listings}
\usepackage{float}
\usepackage{soul}
\usepackage{adjustbox}
\usepackage{colortbl}
\newtheorem{definition}{Definition}[section] 
\usepackage{blindtext,nameref}

\usepackage{enumitem}

\allowdisplaybreaks
\marginsize{2.5cm}{2.5cm}{2.5cm}{2.5cm}

\date{}
\begin{document}
	\begin{frontmatter}
		\title{\textbf{Interval-Valued Fuzzy Fault Tree Analysis through Qualitative Data Processing and its Applications in Marine Operations}}	
		\author{Hitesh Khungla, Kulbir Singh, Mohit Kumar}
		
		\address{Department of Basic Sciences,\\ Institute of Infrastructure, Technology, Research And Management, Ahmedabad, Gujarat-380026, India\\
			Email: hitesh.khungla.iitram@gmail.com, kulbir.singh.17pm@iitram.ac.in, mohitkumar@iitram.ac.in}
		
    \begin{abstract}
    Marine accidents highlight the crucial need for human safety. They result in loss of life, environmental harm, and significant economic costs, emphasizing the importance of being proactive and taking precautionary steps. This study aims to identify the root causes of accidents, to develop effective strategies for preventing them. Due to the lack of accurate quantitative data or reliable probability information, we employ qualitative approaches to assess the reliability of complex systems. We collect expert judgments regarding the failure likelihood of each basic event and aggregate those opinions using the Similarity-based Aggregation Method (SAM) to form a collective assessment. In SAM, we convert expert opinions into failure probability using interval-valued triangular fuzzy numbers. Since each expert possesses different knowledge and various levels of experience, we need to assign weights to their opinions to reflect their relative expertise. We employ the Best-Worst Method (BWM) to calculate the weights of each criterion, and then use the weighting scores to determine the weights of each expert. Ranking of basic events according to their criticality is a crucial step, and in this study, we use the FVI measure to prioritize and rank these events according to their criticality level. To demonstrate the effectiveness and validity of our proposed methodology, we apply our method to two case studies: (1) chemical cargo contamination, and (2) the loss of ship steering ability. These case studies serve as examples to illustrate the practicality and utility of our approach in evaluating criticality and assessing risk in complex systems.
    \end{abstract}
  
\end{frontmatter}
\section{Introduction}
The majority of global trade relies on sea transportation, with a considerable volume of cargo being shipped across the world's oceans. Maritime transport routes encompass various waterways, with straits and channels posing the highest risk of ship accidents. Historical accident data shows that a considerable proportion of incidents have occurred in these congested areas, leading to vessel damage, crew endangerment, and supply chain disruptions. Furthermore, such accidents have far-reaching consequences, causing significant economic losses and impacting the global economy. Maritime transportation carries the largest share of chemical cargo transportation, surpassing other modes of transport such as air, railway, road, and pipeline. Given that a significant portion of maritime trade - approximately 40$\%$ involves the transportation of either fossil fuels or chemicals directly derived from fossil fuels. Transporting chemical cargo requires careful planning and adherence to safety regulations to prevent accidents, spills, and environmental damage.

Marine accidents have emerged as a crucial research priority among scholars due to their devastating consequences, including loss of life and significant economic losses that impact the global economy. Various risk analysis methods, including traditional techniques like Hazard and Operability Study (HAZOP) and Failure Modes and Effects Analysis (FMEA)\cite{YAZDI2017113}, have been utilized in the literature to prevent or mitigate accidents. However, in recent times, Fault Tree Analysis (FTA)\cite{watson1961} is a widely employed method for evaluating the reliability of complex systems.

FTA method provides a systematic and structured framework for evaluating the reliability of complex systems using qualitative data as well as quantitative data. FTA uses 'AND' and 'OR' gates to create a graphical relationship between the Top Event (TE) and the Basic Events (BEs). Once the root causes of a particular problem have been identified, FTA is employed to assess the probability of an unwanted incident happening. The FTA decomposes the occurrence of faults into sequential branches, with each stage undergoing analysis until the fundamental causes or boundary conditions are identified. FTA is a powerful to calculate the Failure Probability (FP) of the TE, but it has several limitations, such as dependence on probability data. However, in many complex systems, the FPs of BEs are unknown. In such cases, we use qualitative data to compute the FP of TE. To reduce uncertainty in probability data, Tanaka et. al \cite{Tanaka} propose the use of Fuzzy Extended Fault Tree Analysis (FFTA) to compute the probability of TE using fuzzy possibilities. 

Researchers have extensively employed FFTA in the literature to assess the failure probabilities of systems. FFTA has proven to be a valuable tool for solving reliability assessment challenges in various engineering problems. For example, Wang et al. (2013)\cite{wang2013} applied FFTA to calculate the probability of crude oil tank fires and explosions. In another study, Lavasani et al. (2015)\cite{lavasani} employed FFTA to calculate the failure rates of BEs in the petrochemical process industry. Shi et al. (2018)\cite{SHI2018991} used FFTA to examine the risk of gas and dust explosions in coal mines. Parveen et al. (2019)\cite{PARVEEN2018} applied FFTA to assess the reliability of solar photovoltaic systems in the renewable energy sector. Mottahedi and Ataei (2019)\cite{MOTTAHEDI2019165} applied FFTA to investigate the likelihood of coal bursts in the mining industry. FFTA has been used in many different areas, proving its usefulness and adaptability. The maritime industry frequently adopts FFTA as a preferred method for conducting risk assessments and studies. Mentes and Helvacioglu (2011)\cite{mentes} conducted a risk analysis of spread mooring systems using FFTA. The study relied on expert opinions to determine the probability values of BEs. Senol et al. (2015)\cite{SENOL} employed FFTA to identify the root causes of cargo contamination and quantify its probability. M. Yazdi and S. Kabir employed FFTA and Bayesian network-based approach for risk analysis in process industries\cite{YAZDI17}\cite{YAZDI2017507}. Kuzu et al. (2019)\cite{KUZU2019} conducted a risk assessment of ship mooring operations using FFTA, identifying potential hazards and evaluating their likelihood and impact. S. Gurgen et al. (2023)\cite{GURGEN2023114419} applied FFTA to compute the probability of the event "Loss of Ship Steering Ability". M. Kumar and M. Kaushik (2022)\cite{KAUSHIK2022103229}\cite{KAUSHIK2023113411} assessed the failure probability of oil and subsea production systems using FFTA in an intuitionistic fuzzy environment and evaluated risk analysis of ship mooring operations using FFTA and bayesian network in an intuitionistic fuzzy environment.

Within fuzzy set theory\cite{zadeh}, the degree of membership for a fuzzy number, representing the failure possibilities of a BE is typically characterized by a single value. However, there are situations where experts may harbor uncertainty regarding these membership values. Therefore, it is often more appropriate to represent the membership values of BE using intervals of possible real numbers rather than specific, precise real numbers. Ching-Fen Fuh et al. \cite{ching} used ($\lambda$, 1) interval-valued fuzzy numbers to evaluate fuzzy reliability of systems. P. Kumar and S.B.Singh \cite{kumarfuzzy} used FFTA using Level ($\lambda$,$\rho$) Interval-valued fuzzy numbers to evaluate fuzzy reliability of systems. 

To compute the probability of the TE using qualitative data, aggregation of expert opinions is a crucial step. In the literature, there are several techniques to aggregate expert's opinions. For example similarity aggregation method \cite{hsu}, linear opinion pool \cite{clemen}, arithmetic averaging operation, fuzzy analytic hierarchy process \cite{SAHIN}, and distance-based aggregation method \cite{kumar}. Additionally, when aggregating expert opinions, assigning weights to each expert is important as it ensures that their opinions and judgments are properly represented and factored into the overall evaluation. Various techniques exist to compute expert weights, such as BWM\cite{REZAEI2016126}, AHP\cite{HELLENDOORN1993}, FLLSM\cite{van}, fuzzy programming method\cite{mikhailov}, FAHP\cite{MIKHAILOV03}, WS method\cite{miri}, etc.

Many studies have looked at risks in different areas of the maritime industry. However, there are very few studies on the risks related to ship steering systems. Goksu et al. (2023)\cite{GOKSU2023} evaluated the risks of ship steering gear systems by analyzing accident reports related to steering gear failures, using a fuzzy-bayesian networks approach. Previous research in FFTA has primarily utilized fuzzy numbers or intuitionistic fuzzy numbers, but no notable studies have employed interval-valued fuzzy numbers. This study addresses this gap by applying Interval-Valued Fuzzy Fault Tree Analysis (IVFFTA) for qualitative data using the Similarity Aggregation Method (SAM) to aggregate expert's opinions. Additionally, we will employ the Best-Worst Method (BWM) to compute the weights of the criteria and WS to determine the weights of the experts. To test the applicability of our proposed methodology, we have conducted two case studies: one on chemical cargo contamination and another on loss of ship steering ability. These case studies showcase the versatility and practicality of our approach to evaluating risks in various maritime contexts.

The remainder of this paper is organized as follows: Section 2 introduces fundamental concepts of fuzzy set theory. Section 3 presents our proposed methodology, followed by Section 4, which illustrates its application through two case studies. Finally, Section 5 concludes the paper, summarizing our key findings.

\section{Background}
This section provides an overview of fundamental concepts and definitions of fuzzy set theory.
\begin{definition}
Fuzzy Set \cite{Klir2015}\cite{zadeh} 
\end{definition}
A fuzzy set $\tilde{A}$ defined on universal set $\mathbb{X}$, is an ordered pair $\tilde{A} = \left\lbrace <x, \mu_{\tilde{A}}(x)> : x \in \mathbb{X} \right\rbrace $ where $\mu_{\tilde{A}}(x) : \mathbb{X} \rightarrow  [0,1] $ is membership function of set $\tilde{A}$. A fuzzy set $\tilde{A}$ defined on $\mathbb{R}$ is called a fuzzy number if it possesses the following conditions. \\
(1). $\tilde{A}$ is normal, i.e. $\exists x \in \mathbb{R} \text{ s.t. } \mu_{\tilde{A}}(x)=1.$ \\
(2). $\tilde{A}$ is convex, i.e. $\forall x_1,x_2 \in \mathbb{R}, 0 \le \lambda \le 1$ 
    \begin{equation*}
        \mu_{\tilde{A}}(\lambda x_1 + (1-\lambda)x_2) \ge min\{\mu_{\tilde{A}}(x_1),\mu_{\tilde{A}}(x_2)\}.
    \end{equation*} \\
(3). supp$(\tilde{A})=\{ x \in \mathbb{R}: \mu_{\tilde{A}}(x)>0 \}$ is bounded. \\

Let $ \tilde{A} = \left\lbrace < x, \mu_{\tilde{A}}(x) > \ : x \in \mathbb{R} \right\rbrace $ having a membership function $\mu_{\tilde{A}}(x)$ which is defined as follows:
\begin{equation*}
	\mu_{\tilde{A}}(x) = 
	\left\{\begin{matrix}
		\frac{x-a}{b-a}, \ \ \ \ a \leq x \leq b\\
		\\
            \frac{c-x}{c-b}, \ \ \ \  b \leq x \leq c\\
		\\
		 \ \ 0,\ \ \ \ \ \ \     \text{otherwise}
	\end{matrix}\right.
\end{equation*}
then $\tilde{A}$ is called the Triangular Fuzzy Number (TFN) and it is denoted by $\tilde{A}=(a,b,c)$.

\begin{definition}
    Interval-valued Triangular Fuzzy Number \cite{Zimmermann}
\end{definition}
An Interval-Valued Triangular Fuzzy Number (IVTFN) $\tilde{A}$ on $\mathbb{R}$ is given by \\
$ \tilde{A} = \left\lbrace < x, [\mu_{\tilde{A}^L}(x),\mu_{\tilde{A}^U}(x)] >  \ : x \in \mathbb{R} \right\rbrace, $ where $\mu_{\tilde{A}^L}(x)$ and $\mu_{\tilde{A}^U}(x)$ are defined as follows:
\begin{equation*}
    \mu_{\tilde{A}^L}(x)= \left\{\begin{matrix}
		\frac{x-a}{b-a}, \ \ \ \ a \leq x \leq b\\
		\\
            \frac{c-x}{c-b}, \ \ \ \  b \leq x \leq c\\
		\\
		 \ \ 0,\ \ \ \ \ \ \     \text{otherwise}
	\end{matrix}\right.
\end{equation*}
 
\begin{equation*}
 \mu_{\tilde{A}^U}(x) =  \left\{\begin{matrix}
		\frac{x-e}{b-e}, \ \ \ \ e \leq x \leq b\\
		\\
            \frac{h-x}{h-b}, \ \ \ \  b \leq x \leq h\\
		\\
		 \ \ 0,\ \ \ \ \ \ \     \text{otherwise}
	\end{matrix}\right.
\end{equation*}
 and it is denoted as 
$\mu_{\tilde{A}}(x) = [\mu_{\tilde{A}^L}(x),\mu_{\tilde{A}^U}(x)], x \in \mathbb{R}$ or $\tilde{A} = [\tilde{A}^L,\tilde{A}^U]$.

\begin{figure}[H]
    \centering
    \includegraphics[scale=0.35]{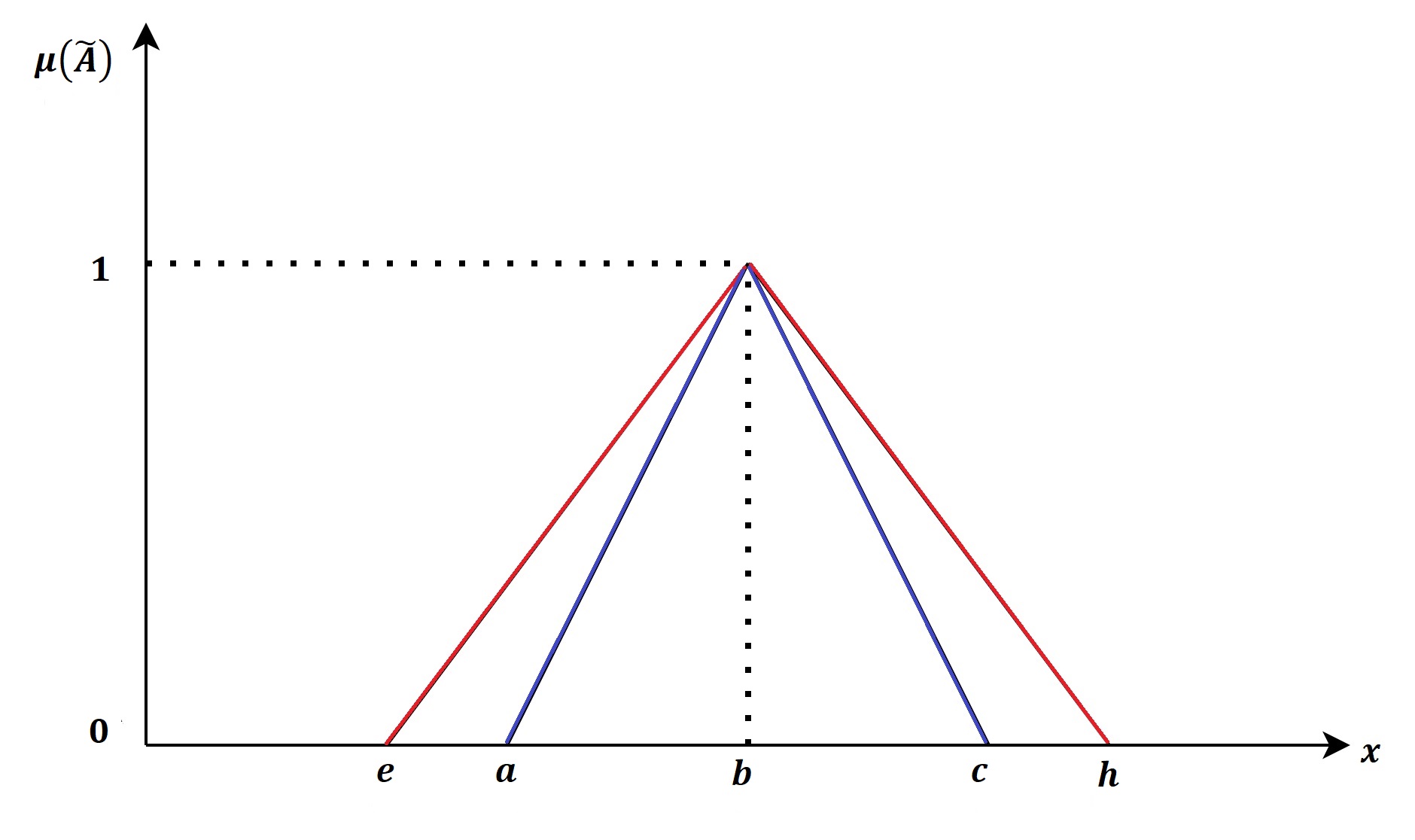}
    \caption{Visual representation of an IVTFN}
    \label{fig1}
\end{figure}

Let $\tilde{A}^L = (a,b,c)$ and $\tilde{A}^U = (e,b,h)$, where $e<a<b<c<h$.
Then the IVTFN is expressed as 
\begin{equation*}
    \tilde{A}= [(a,b,c),(e,b,h)].
\end{equation*}
Based on that, the visual representation of an IVTFN is provided in Fig. \ref{fig1}.

\begin{definition}
    Fuzzy Arithmetic Operations for IVTFN \cite{STANU2016}
\end{definition}
The fuzzy arithmetic operations for an IVTFN are defined as follows:\\
Let $\tilde{A}=[(a_1,b_1,c_1),(e_1,b_1,h_1)]$, $\tilde{B}=[(a_2,b_2,c_2),(e_2,b_2,h_2)]$ be two IVTFN and $k$ be any positive scalar then,\\
(1)
    $\tilde{A} + \tilde{B} = [(a_1+a_2,b_1+b_2,c_1+c_2),(e_1+e_2,b_1+b_2,h_1+h_2)]$ \\
(2) 
    $\tilde{A} - \tilde{B} = [(|a_1-c_2|,|b_1-b_2|,|c_1-a_2|),(|e_1-h_2|,|b_1-b_2|,|h_1-e_2|)]$ \\
(3)
    $\tilde{A} \times \tilde{B} = [(a_1 \times a_2,b_1 \times b_2,c_1 \times c_2),(e_1 \times e_2,b_1 \times b_2,h_1 \times h_2)]$ \\
(4) 
    $\tilde{A} \div \tilde{B} = [(a_1 \div c_2,b_1 \div b_2,c_1 \div a_2),(e_1 \div h_2,b_1 \div b_2,h_1 \div e_2)]$ \\
(5)
    $k\tilde{A}=[(ka_1,kb_1,kc_1),(ke_1,kb_1,kh_1)]$

\section{Proposed Methodology}
In this section, we will aggregate expert opinions using the SAM and calculate the weights of each expert using the BWM and weighting scores. We will calculate the FP of TE and compute the FVI measure to assess the criticality of BEs.

\begin{enumerate}
\item \textit{Experts judgments}

This step generates qualitative possibilities for all BEs through expert elicitation. The experts involved in the process possess a deep understanding of the system's environment and hold valuable knowledge and experience related to the system's processes. In assessing component failure likelihood, experts commonly use natural language terms, such as 'Low' or 'High', instead of assigning exact numerical probabilities. Each expert selects a linguistic term from a predefined set of terms to respond to specific questions concerning the likelihood of each BE. In this paper, seven linguistic terms: Very Low (VL), Low (L), Medium Low (ML), Medium (M), Medium High (MH), High (H) and Very High (VH) are utilized to assess the possibilities of BEs. The likelihood of system failure is impacted by the probability of each BE, which is evaluated based on the expertise and experience of knowledgeable professionals. Each expert's opinion on how likely a part is to fail may be different, therefore we must weigh each expert's judgment. Here, BWM is used to determine the weights of criteria and after that, we calculate the weights of the experts by multiplying these criteria weights with the weighting score of each criterion for each expert. The steps of the BWM are described as follows:
\begin{enumerate}
    \item \textit{Identify and categorize decision criteria:}\\
    We choose a set of decision criteria, denoted as $(c_1,c_2,c_3,...,c_n)$, and assign weights to each of these criteria.

    \item \textit{Determine the best and worst criteria:}\\
    From the set of criteria $(c_1,c_2,c_3,...,c_n)$, The criteria that are considered the best and worst are chosen and labeled as $c_B$ (best criteria) and $c_W$ (worst criteria).
   
    \item \textit{Determine criteria preferences over best and worst criterion:}\\
     A value between 1 and 9 is assigned to express the degree of preference for the most important criteria compared to the rest and the same for the worst criteria. This will result in a vector $A_b = (a_{b1},a_{b2},a_{b3},...,a_{bn})$, where $a_{bj}$ represents the preference of the best criteria over criteria $j$ and $A_w = (a_{1w},a_{2w},a_{3w},...,a_{nw})$, where $a_{jw}$ represents the preference of the $j^{th}$ criteria over the worst criteria. The vectors $A_b$ and $A_w$ are jointly known as a pairwise comparison system, denoted as $(A_b,A_w)$. 

    \item \textit{Calculate Optimal Weights:}\\
The weight set is obtained by solving the system of equations $\frac{w_b}{w_j}=a_{bj}$ and $\frac{w_j}{w_w}=a_{jw}$. However, in cases where the system has no solution, The optimal weight set is obtained by solving the optimization problem that minimizes the maximum deviation between the weight ratios and the given comparison values. For that consider the following optimization problem:

    \begin{equation} \label{bwm}
    \begin{split}
        &\text{min} \ \xi \\
        &\text{sub. to:}\quad \vert \frac{w_b}{w_j}-a_{bj} \vert \le \xi \ \ \  \text{and} \ \ \ \vert \frac{w_j}{w_w}-a_{jw} \vert \le \xi \ \ \ \text{for all } j. \\
        &\quad \quad\quad \quad \sum_{j}{w_j}=1, w_j \ge 0 \ \ \text{for all } j.
        \end{split}
    \end{equation}
    By solving problem \ref{bwm}, we obtain the optimal weights $(w_1^*,w_2^*,w_3^*,...,w_n^*)$ and $\xi^*$ 

    \item \textit{Calculate the consistency ratio based on the consistency index:}\\
    The consistency ratio is determined by dividing the optimal value $\xi^*$ by the consistency index, as illustrated below:
    \begin{equation*}
        \text{Consistency Ratio}=\frac{\xi^*}{\text{Consistency Index}},
    \end{equation*}
    where Consistency Index$=sup\{\xi^*:\xi^* $ is the optimal objective value of problem \ref{bwm} for some $(A_b,A_w)$ having the given $a_{bw}$\}. The consistency ratio value closer to zero represents a higher degree of consistency.
\end{enumerate}

The set of optimal weights $(w_1^*,w_2^*,w_3^*,...,w_n^*)$ signifies the relative significance of the criteria $(c_1,c_2,c_3,...,c_n)$. We determine the experts' weights by multiplying these criteria weights with the weighting scores of each criterion for each expert. Weighting scores for each expert can be calculated using Table 1 and Table 2.

\begin{figure}[H]
    \centering
    \includegraphics[width=1\linewidth]{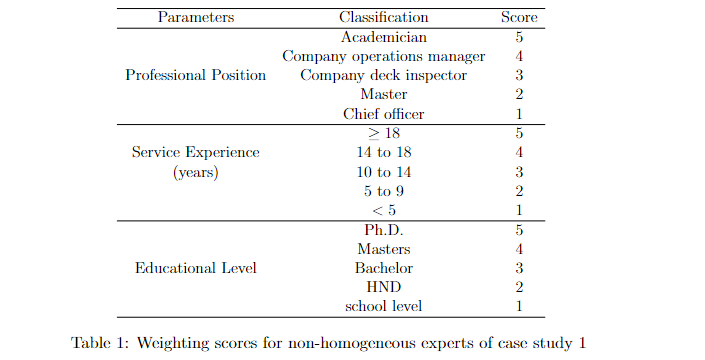}
   \end{figure}

\begin{figure}[H]
    \centering
    \includegraphics[width=1\linewidth]{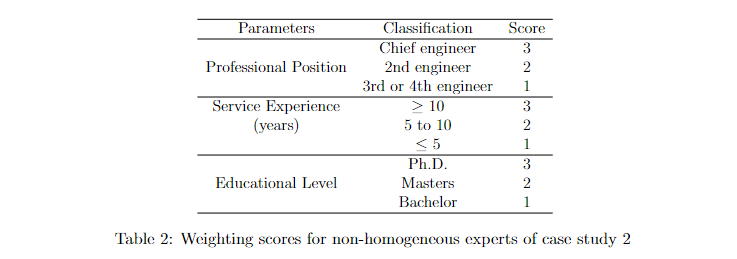}
    \end{figure}

\item \textit{Fuzzification of experts judgments using IVTFN} \\
Experts may offer varying judgments based on their unique experiences and knowledge, resulting in differing opinions. The challenge lies in consolidating all these judgments and reaching a conclusion. Aggregation can be performed using a simple average, but this approach neglects the significance of expertise and experience. A weighted average is a more effective way to combine opinions, as it gives more weight to the insights of highly experienced experts with high scores while lessening the impact of less experienced experts with lower scores. The Similarity Aggregation Method (SAM)\cite{KABIR} not only prioritizes consensus but also assigns significance to similar opinions. The SAM involves the following steps:
\begin{enumerate}
    \item \textit{Degree of similarity} \\
    Let $\tilde{A}=[(a_1,b_1,c_1),(e_1,b_1,h_1)] \ \text{and} \ \tilde{B}=[(a_2,b_2,c_2),(e_2,b_2,h_2)]$, then similarity between this two numbers is defined as:\cite{Similarity}
    \begin{align*}
        S(\tilde{A},\tilde{B}) &= \sqrt{S((a_1,b_1,c_1),(a_2,b_2,c_2)) \times S((e_1,b_1,h_1),(e_2,b_2,h_2))} \\
        &= \sqrt{\bigg(1-\frac{|a_1-a_2|+|b_1-b_2|+|c_1-c_2|}{3}\bigg) \bigg(1-\frac{|e_1-e_2|+|b_1-b_2|+|h_1-h_2|}{3}\bigg)}
    \end{align*}
    \item \textit{Average Agreement degree (AA($E_u$))} \\
    Let $(E_1,E_2,...,E_u,...E_n)$ be set of n experts then the Average Agreement degree for each expert $E_u$ is defined as:
    \begin{equation*}
        AA(E_u)=\frac{1}{n-1}\sum_{v=1 , u \neq v}^{n}S(\tilde{R_u},\tilde{R_v}),
    \end{equation*}
    where $\tilde{R_u}$ and $\tilde{R_v}$ is opinions of expert $E_u$ and $E_v$ respectively.
    \item \textit{Relative Agreement degree (RA($E_u$))} \\
    Among the set of n experts Relative Agreement degree for expert $E_u$ is defined as:
    \begin{equation*}
        RA(E_u)=\frac{AA(E_u)}{\sum_{u=1}^n AA(E_u)}
    \end{equation*}
    \item \textit{Consensus Coefficient degree (CC($E_u$))} \\
    \begin{equation*}
        CC(E_u)=\beta w(E_u) + (1-\beta) RA(E_u)
    \end{equation*}
    where $\beta$ is a relaxation factor and $(0 \leq \beta \leq 1 )$ indicates the importance of $w(E_u)$ over $RA(E_u)$. A higher value of $\beta$ assigns more importance to expert opinions, while a lower value favors SAM outputs. Typically, we set $\beta = 0.5$, which assigns equal weights to both expert opinions and SAM outputs.
    
    \item \textit{Aggregation of experts' opinion, $\tilde{R}_{AG}$} \\
    \begin{equation} 
    \tilde{R}_{AG}=\prod_{u=1}^{n}CC(E_u)\times\tilde{R_u}.
    \end{equation}
\end{enumerate}
    \item Defuzzification process \\
   Defuzzification is a technique used to transform a fuzzy number into a precise, crisp value. Various methods for defuzzification have been proposed in the literature, including the Center of Area (COA) method, Center of Sums (COS) method, Center of Maximum (COM) method, and others\cite{1993defuzzification} \cite{FORTEMPS1996} \cite{LEEKWIJCK1999} \cite{OPRICOVIC2003}.
    For an IVTFN $\tilde{A}=[(a,b,c),(e,b,h)]$ we have,
    \begin{equation*}
        A^*= \frac{4b+a+c+e+h}{8}
    \end{equation*}
    \item Transforming Crisp Failure Possibility (CFP) into Failure Probability (FP) \\The function introduced by Onisawa (1988)\cite{ONISAWA198887} has been widely adopted by numerous scholars to convert CFP into FP in fuzzy quantitative risk. analysis. Which is defined as follows:
    \begin{equation*}
        FP=\begin{cases}
    \frac{1}{10^K}, & \text{when CFP $\neq$ 0}\\
    0, & \text{when CFP=0}
  \end{cases}
    \end{equation*}
    where,
    \begin{equation*}
        K=\bigg(\frac{1-CFP}{CFP}\bigg)^{\frac{1}{3}}\times 2.301
    \end{equation*}

    However, the Onisawa function does not work for all areas. Yu et al. (2022)\cite{YU2022} proposed an improved methodology to convert CFP into FP. Which is defined as follows:
    \begin{equation*}
        K=\begin{cases}
   -0.72ln(CFP) + 2.839 & \text{$0 \le CFP \le 0.2$}\\
    4.523 - 3.287CFP & \text{$0.2 \le CFP \le 0.8$}\\
    3.705 \bigg(\frac{1-CFP}{CFP}\bigg)^{0.445} & \text{$0.8 \le CFP \le 1$} 
  \end{cases}
  \end{equation*}
    \item Calculating top event probability P(TE) \\
    Let $q_i(t)$ denote the probability that the basic event $BE_i$ occurs at time $t$. Similarly, let $P(TE)$ represent the probability that the top event occurs at time $t$.

    If all BEs $(BE_1,BE_2,BE_3,...,BE_n)$ are connected with a single AND gate then the probability of TE is determined as:\cite{hoyland}
    \begin{equation*}
        P(TE)=\prod_{i=1}^{n}q_i(t).
    \end{equation*}
     If all BEs $(BE_1,BE_2,BE_3,...,BE_n)$ are connected with a single OR gate then the probability of TE is determined as:
    \begin{equation*}
        P(TE)=1-\prod_{i=1}^{n}(1-q_i(t)).
    \end{equation*}

\item Importance analysis \\
 Importance analysis is a technique applied to quantify the risk associated with a particular event and to determine the relative importance of BEs in a complex system, establishing a ranking to guide further analysis or mitigation efforts. This ranking is crucial for identifying the weak links in the system in terms of reliability and understanding the contribution of each BE to the system's overall reliability. In our study, we have used the FVI measure to rank the importance of BEs. For the $BE_i$, where $i=1,2,3,...,n$, the FVI measure is calculated as follows:
 \begin{equation*}
     FVI(BE_i)=\frac{P_{TE}-P_{TE}(BE_i=0)}{P_{TE}}
 \end{equation*}
\end{enumerate}

\section{Case Study}
In this section, we will demonstrate the practical applicability of our proposed methodology by applying it to two real-world case studies: 'Chemical Cargo Contamination' and 'Loss of Ship Steering Ability'. These examples will illustrate the effectiveness of our methodology in addressing complex maritime safety issues.

\subsection{Case Study 1: Chemical Cargo Contamination}
\begin{figure}[H]
    \centering
    \includegraphics[width=13cm,height=10cm,keepaspectratio]{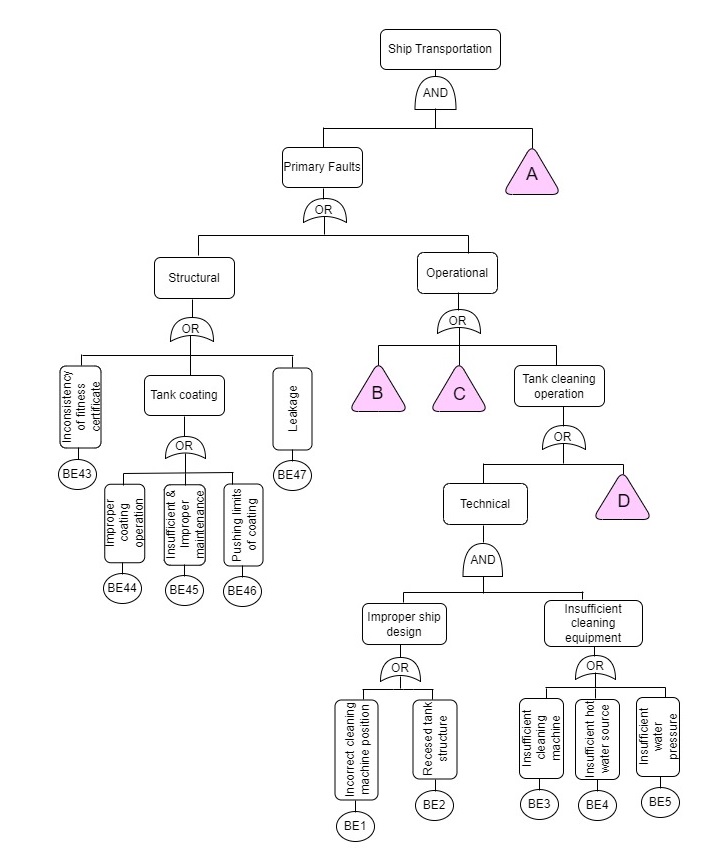}
    \caption{Fault tree of top event \textbf{chemical cargo contamination}}
    \label{f1}
\end{figure}

"Cargo contamination on board" is the TE for this case study. Here we are using a fault tree which was developed by Senol et. al.\cite{SENOL}. In this article, the IVFFTA technique is employed for a set of non-homogeneous experts. Structured fault trees are indicated in Fig \ref{f1} to Fig \ref{f5}.

\begin{figure}[H]
    \centering
    \includegraphics[width=13cm,height=10cm,keepaspectratio]{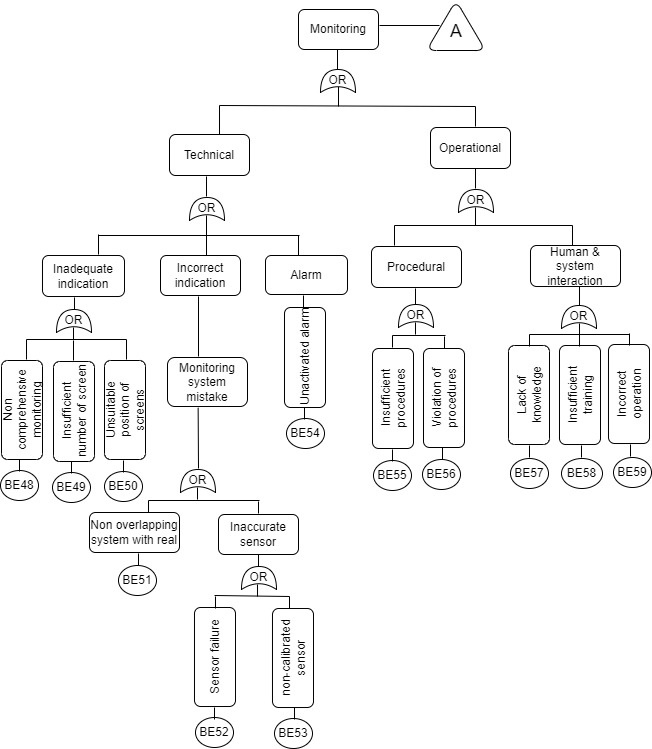}
    \caption{Fault tree for event \textbf{A} of chemical cargo contamination}
    \label{f2}
\end{figure}

\begin{figure}[H]
    \centering
    \includegraphics[width=13cm,height=10cm,keepaspectratio]{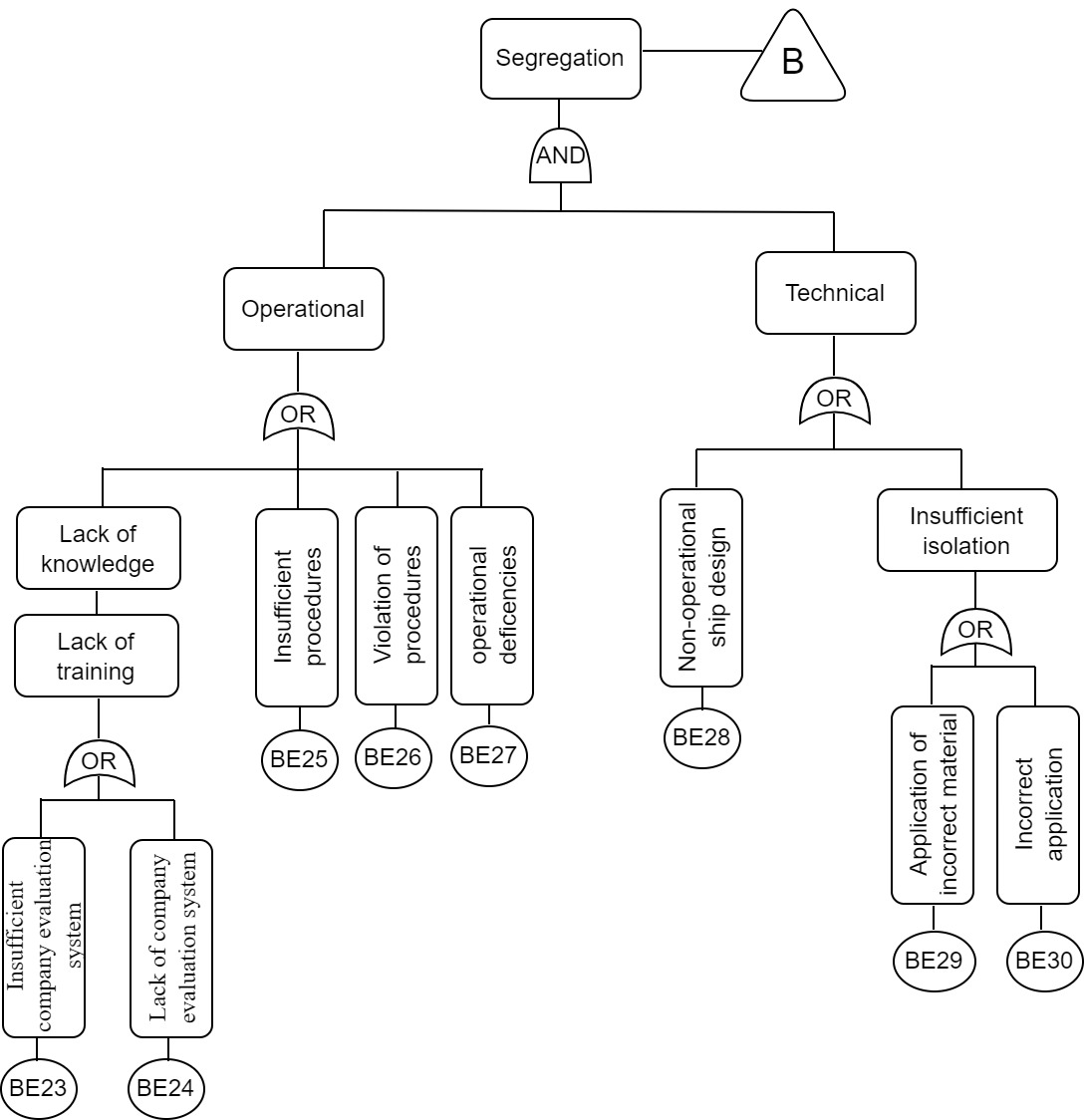}
    \caption{Fault tree for event \textbf{B} of chemical cargo contamination}
    \label{f3}
\end{figure}

\begin{figure}[H]
    \centering
    \includegraphics[width=13cm,height=10cm,keepaspectratio]{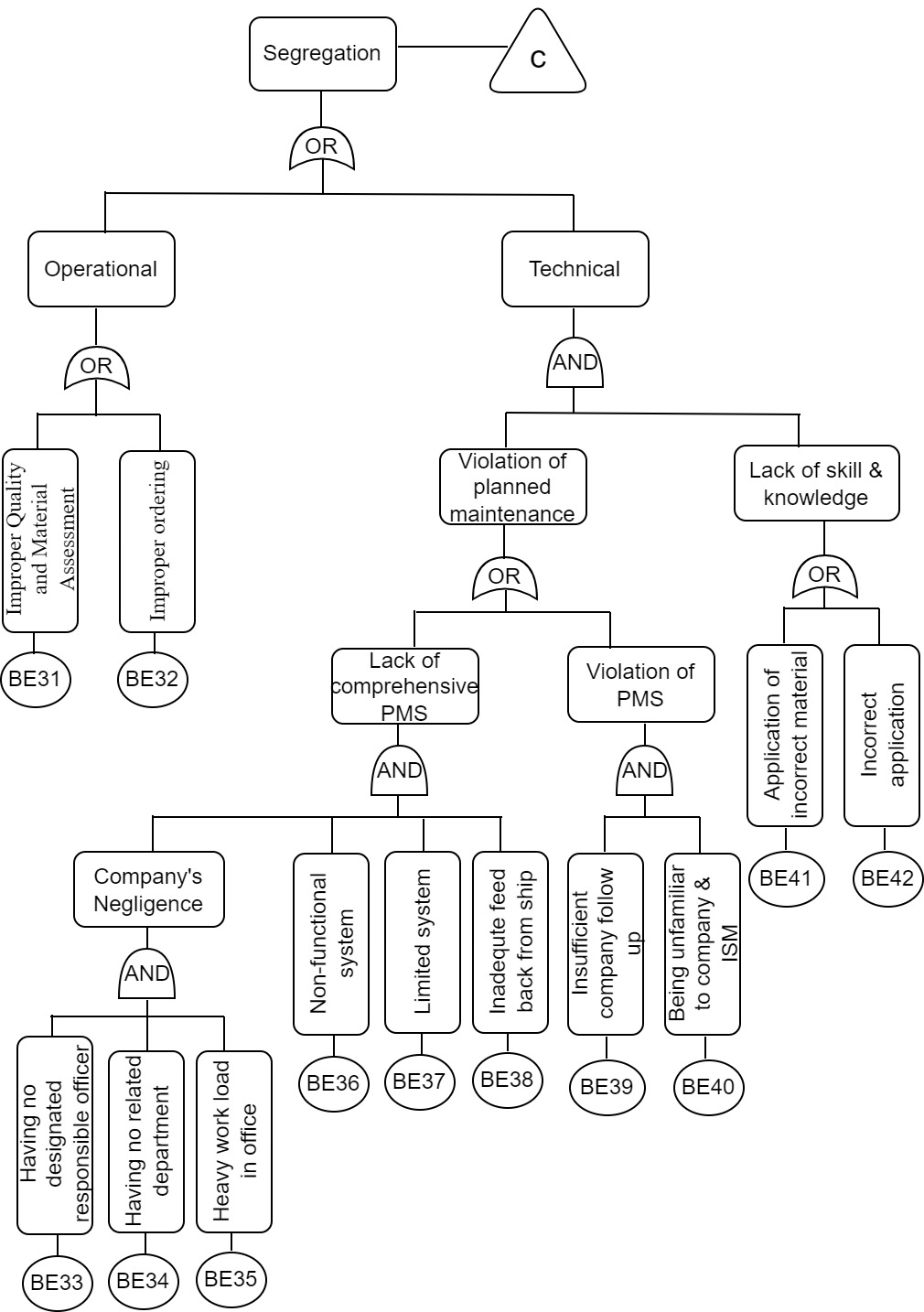}
    \caption{Fault tree for event \textbf{C} of chemical cargo contamination}
    \label{f4}
\end{figure}

\begin{figure}[H]
    \centering
    \includegraphics[width=13cm,height=10cm,keepaspectratio]{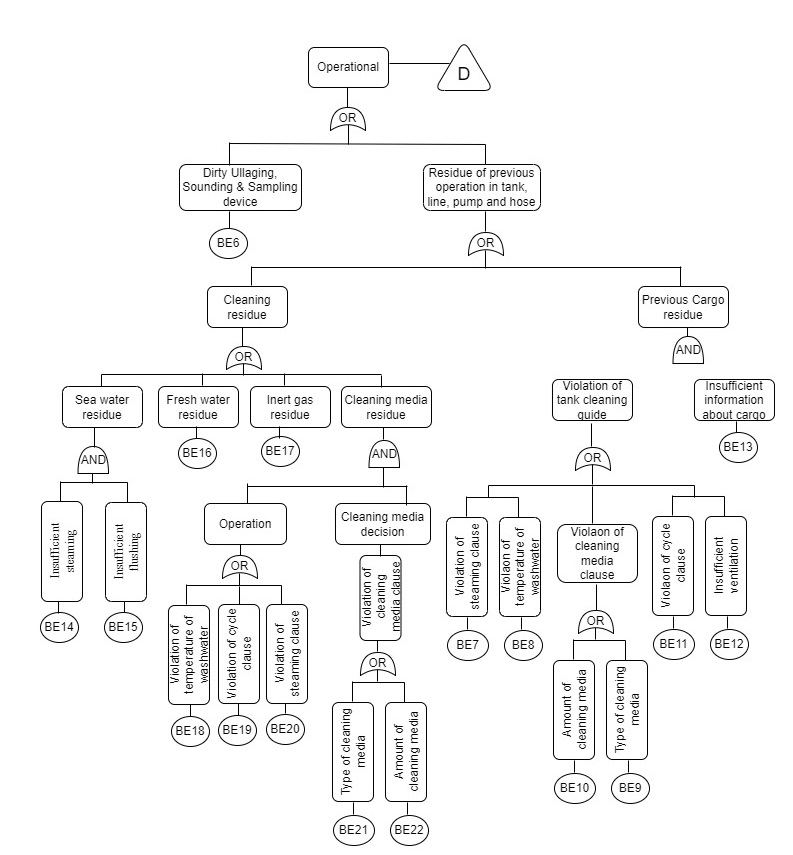}
    \caption{Fault tree for event \textbf{D} of chemical cargo contamination}
    \label{f5}
\end{figure}

\begin{enumerate}
    \item \textit{Experts Judgements} \\
    To evaluate the possibilities of BEs, a group of three experts is chosen. Experts' opinions about the likelihood of BEs occurring can be subjective due to variations in their levels of knowledge and experience. Therefore, to signify the comparative quality of different experts, weights are calculated using BWM and weighting scores using Table 1 and shown in Table 5. In this study, seven linguistic terms have been chosen for the subjective assessment of the probabilities of BEs. which is given in Table 3. The opinions of experts for all BEs are given in Table 4.

\begin{figure}[H]
    \centering
    \includegraphics[width=1\linewidth]{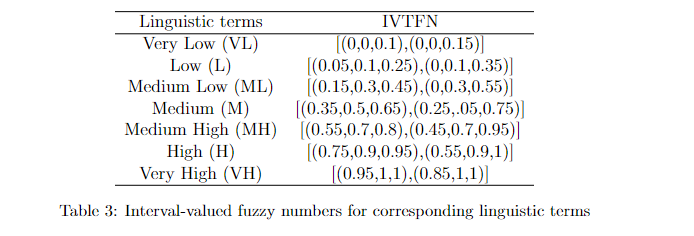}
    \end{figure}

\begin{figure}[H]
    \centering
    \includegraphics[width=1\linewidth]{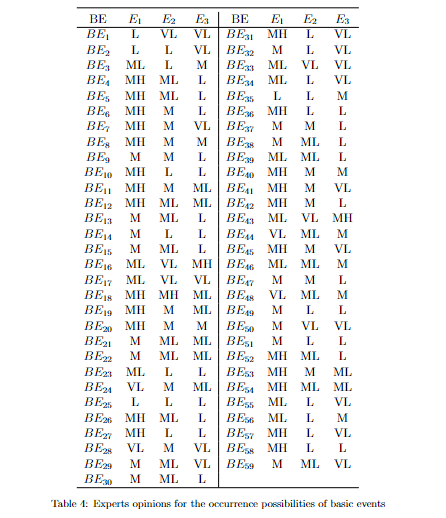}
    \end{figure}

\begin{figure}[H]
    \centering
    \includegraphics[width=1\linewidth]{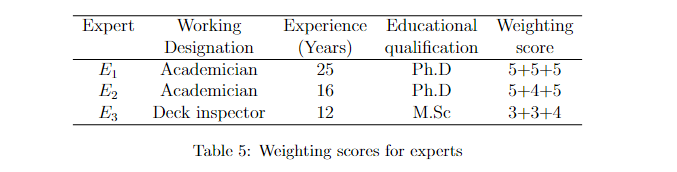}
    \end{figure}

To assign weights to the experts, three criteria designation ($c_1$), Experience ($c_2$), and qualification ($c_3$) are selected. The decision-maker determines the preference of these three criteria. Assume that he chooses experience as the best criterion, and designation as the worst criterion. The preferences of these three criteria by decision-maker are listed in Table 6 and weights of criteria are calculated using BWM by solving a following optimization problem.

\begin{figure}[H]
    \centering
    \includegraphics[width=1\linewidth]{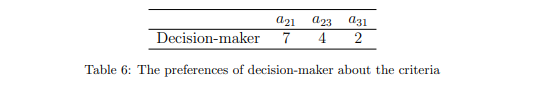}
    \end{figure}

Here, $a_{ij}$ refers the preference of criteria $i$ over criteria $j$.\\

The optimization problem can be formulated as \\
\begin{equation*}
    \begin{split}
    &\text{min} \ \ \xi \\
    &\text{sub. to:} \ \vert \frac{w_2}{w_1}-7 \vert \le \xi,\ \vert \frac{w_2}{w_3}-4 \vert \le \xi,\ \vert \frac{w_3}{w_1}-2 \vert \le \xi, \\
    &\sum_{j=1}^{3} w_j = 1 \ \ \text{and} \ \ w_j \ge 0 \ \ \text{for } j=1,2,3.
    \end{split}
\end{equation*}

By solving the above optimization problem, weights for the criteria $(c_1,c_2,c_3)$ are (0.1000, 0.7146, 0.1854) and the resulting consistency ratio is 0.0391. This consistency ratio indicates that the results are consistent since its value is close to 0. Using weights of criteria and WSs of each expert, weights of the expert can be calculated as follows.

\begin{equation*}
    \begin{bmatrix}
    w_1 \\ w_2 \\ w_3
    \end{bmatrix}
    =
    \begin{bmatrix}
         & WS(c_1) & WS(c_2) & WS(c_3) \\
     E_1 & 5 & 5 & 5 \\
     E_2 & 5 & 4 & 5 \\
     E_3 & 3 & 3 & 4 
    \end{bmatrix}
    \times
    \begin{bmatrix}
        0.1000 \\ 0.7146 \\ 0.1854
    \end{bmatrix}
\end{equation*}

Based on the calculations the normalized weights of experts $(w_1,w_2,w_3)$ are (0.4042, 0.3383, 0.2575). \\

\item Fuzzification of expert's judgments using IVTFN \\
At this stage, the individual choices of the three experts are aggregated using the SAM approach, outlined in the methodology section. For obtaining the consensus coefficient, the value of $\beta$ is taken as 0.5. Aggregation process for $BE_{53}$ is shown in Table 7 to Table 11.

In Table 7, Expert's opinions are given in IVTFN using that we can find the similarity function value, average agreement value, relative agreement value, and consensus coefficient of each experts which is given in Table 8 to Table 11.

\begin{figure}[H]
    \centering
    \includegraphics[width=1\linewidth]{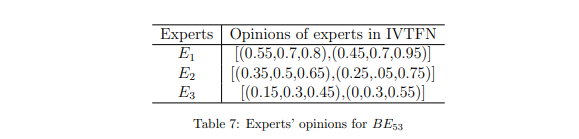}
\end{figure}

\begin{figure}[H]
    \centering
    \includegraphics[width=1\linewidth]{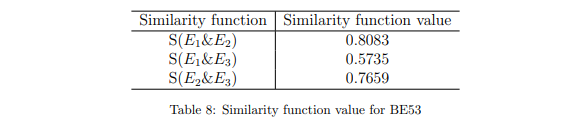}
\end{figure}

\begin{figure}[H]
    \centering
    \includegraphics[width=1\linewidth]{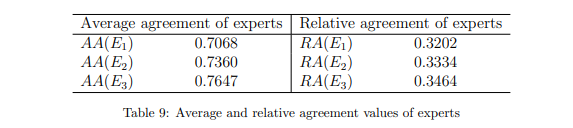}
\end{figure}

\begin{figure}[H]
    \centering
    \includegraphics[width=1\linewidth]{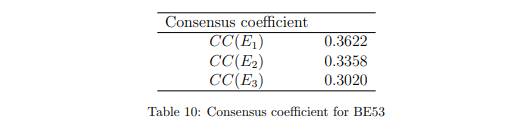}
\end{figure}

\begin{figure}[H]
    \centering
    \includegraphics[width=1\linewidth]{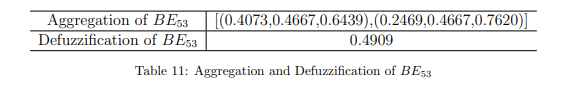}
\end{figure}

\item Defuzzification of BEs \\

Using the defuzzification technique defined as in methodology section we can defuzzify each IVTFN into a crisp value. For example defuzzifued value of $BE_{53}$ is $0.4909.$\\

\item Converting CFP into FP \\

Using the function developed by Yu et al.\cite{YU2022}, we can convert each  CFP into FP. FP of each basic event are given in Table 12.

\begin{figure}[H]
    \centering
    \includegraphics[width=1\linewidth]{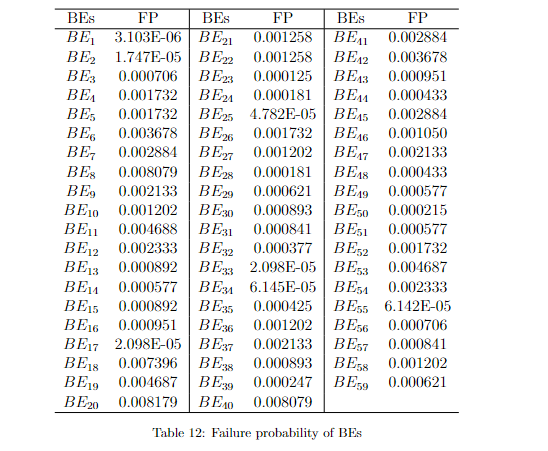}
\end{figure}

\item Calculating top event probability \\
By evaluating the fuzzy probabilities of basic events and the fault tree structure, we have determined the Top Event (TE) probability to be $2.834E-04$, representing the likelihood of the TE occurring. \\

\item Importance analysis \\
The FVI measure serves as an indicator of the contribution of BEs to the probability of the TE. Calculating the FVI measure for each BE enables the prioritization of BEs based on their impact on the probability of the TE. Graph of FVI measure is given in Fig. \ref{fig2}. From this graph, we can conclude that BE53 and BE6 are the most critical events.
\begin{figure}[H]
    \centering
    \includegraphics[width=0.9\linewidth]{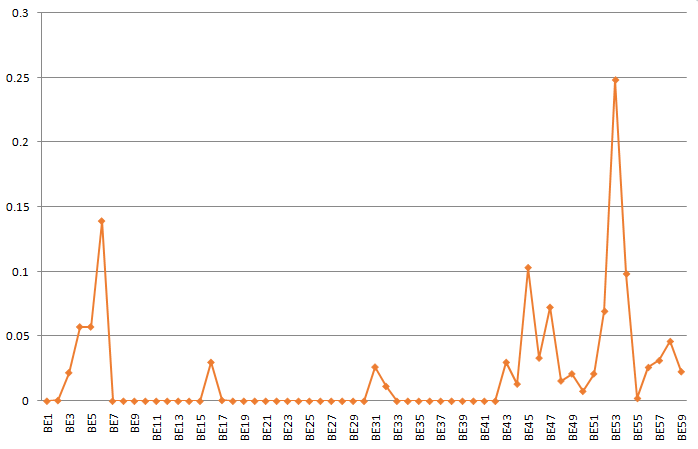}
    \caption{FVI measure of each BE}
    \label{fig2}
\end{figure}
\end{enumerate}

\subsection{Results and discussion}
Using the proposed approach, FP of TE after applying IVFFTA is $2.834E-04$. If a company has 10 chemical tankers, each conducting 6 cargo operations monthly, resulting in a total of 60 cargo operations per month. It means the company makes 720 cargo operations per year. using FP of TE we can calculate the period of occurrence is approximately 4 years and 10 months. According to Senol et. al \cite{SENOL}, they have observed that either major or minor cargo contamination events occur once in approximately four years. As our proposed approach gives it 4 years and 10 months it is very close to the real scenario. 

Furthermore, we have performed a ranking of each BE and presented the results in the following table. Notably, our ranking closely aligns with the ranking provided by Senol et al\cite{SENOL}. This strong correlation provides strong evidence for the accuracy of our method. 

\begin{figure}
    \centering
    \includegraphics[width=1\linewidth]{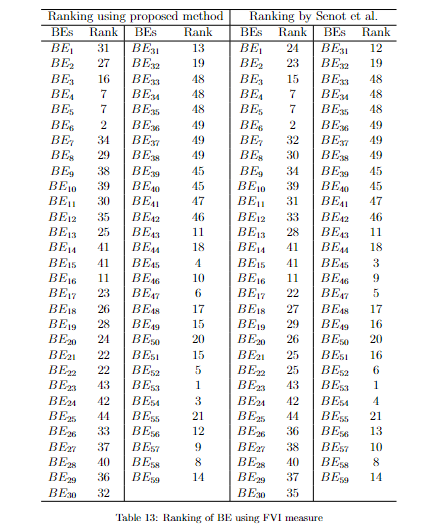}
\end{figure}

\subsection{Case Study 2: Loss of Ship Steering Ability}
\begin{figure}[H]
    \centering
    \includegraphics[scale=0.35]{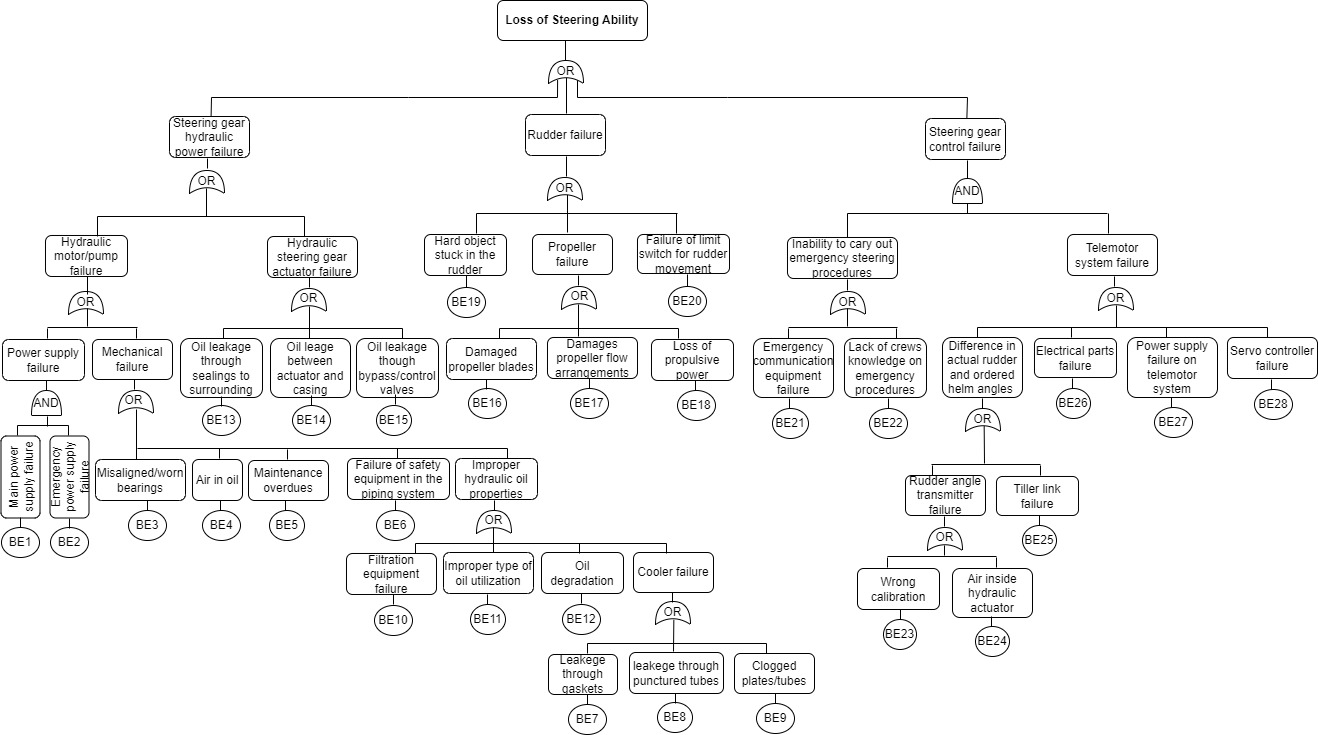}
    \caption{Fault tree of top event loss of ship steering ability}
    \label{c2f1}
\end{figure}

 The three primary contributors to the loss of ship steering ability are the failure of the steering gear's hydraulic power, rudder failure, and failure of the steering gear's control system. Here, we have used a fault tree structure developed by S.Gurgen et al\cite{GURGEN2023114419}. We have employed our algorithm for the set of five non-homogeneous experts. The structure of the fault tree for the 'Loss of Steering Ability' scenario is presented in Fig. \ref{c2f1}.

\begin{enumerate}
    \item Experts Judgements \\
    To calculate the possibilities of BEs, a non-homogeneous group of five experts is chosen. Since each expert has unique service experience and varying levels of education, they inevitably hold different opinions. Therefore, in this case, we will assign weights to their opinions using the BWM and weighting scores. The linguistic terms provided by each expert are presented in Table 14. \\

\begin{figure}[H]
    \centering
    \includegraphics[width=1\linewidth]{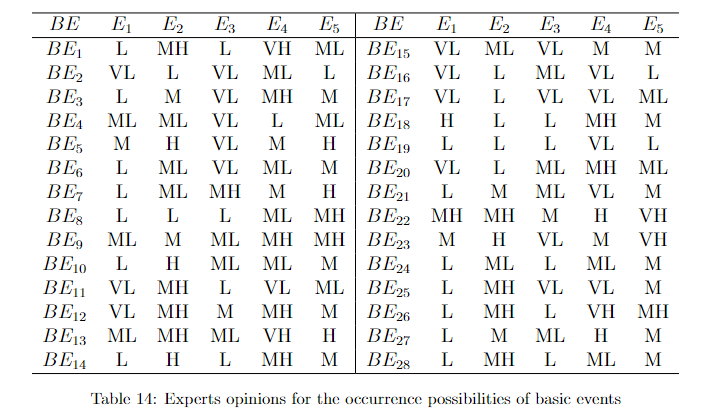}
\end{figure}

Table 15 details the experience, designation, and qualification of each expert, which are used to calculate their respective weighting scores, reflecting their relative expertise and influence on the assessment.

\begin{figure}[H]
    \centering
    \includegraphics[width=1\linewidth]{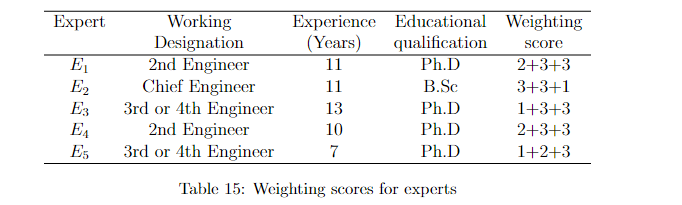}
\end{figure}

Now that we have determined the weights for each criterion, we can proceed to calculate the weights for each expert, which are calculated as follows:

    \begin{equation*}
    \begin{bmatrix}
    w_1 \\ w_2 \\ w_3 \\ w_4 \\ w_5
    \end{bmatrix}
    =
    \begin{bmatrix}
         & WS(c_1) & WS(c_2) & WS(c_3) \\
     E_1 & 3 & 3 & 2 \\
     E_2 & 1 & 3 & 3 \\
     E_3 & 1 & 3 & 3 \\
     E_4 & 3 & 3 & 2 \\
     E_5 & 3 & 2 & 1 \\
    \end{bmatrix}
    \times
    \begin{bmatrix}
        0.1000 \\ 0.7146 \\ 0.1854
    \end{bmatrix}
\end{equation*}

Based on the calculations the normalized weights of experts $(w_1,w_2,w_3,w_4,w_5)$ are $(0.2204,0.2192,0.2059,0.2203,0.1342).$ \\

    \item Fuzzification of experts judgments using IVTFN \\
    This stage involves aggregating the choices of the five experts using the formulas presented in the methodology section. With $\beta$ set to 0.5, the consensus coefficient for each BE is calculated. The aggregation process for BE13 is displayed in Tables 16 to 20.

\begin{figure}[H]
    \centering
    \includegraphics[width=1\linewidth]{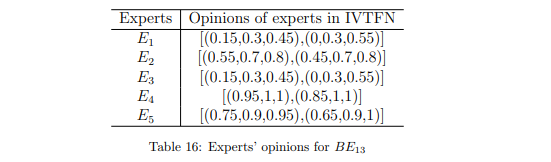}
\end{figure}

\begin{figure}[H]
    \centering
    \includegraphics[width=1\linewidth]{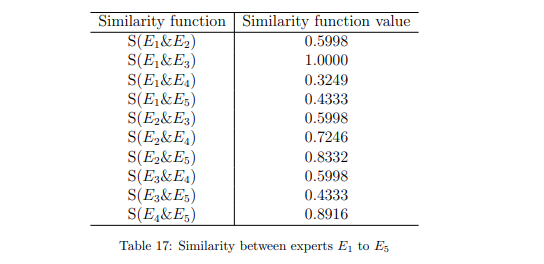}
\end{figure}

\begin{figure}[H]
    \centering
    \includegraphics[width=1\linewidth]{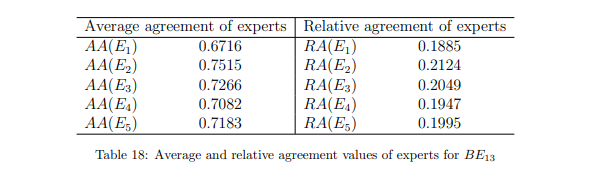}
\end{figure}

\begin{figure}[H]
    \centering
    \includegraphics[width=1\linewidth]{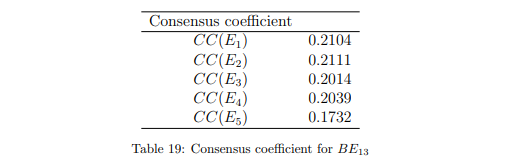}
\end{figure}

\begin{figure}[H]
    \centering
    \includegraphics[width=1\linewidth]{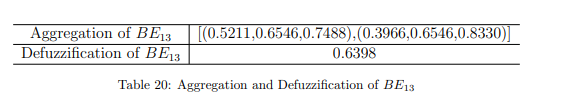}
\end{figure}

    \item Defuzzification of BEs \\
By applying the defuzzification technique outlined in the methodology section, we can convert each IVTFN into a crisp value. For instance, the defuzzified value of $BE_{13}$ is 0.6398.
    
    \item Converting CFP into FP \\
In this step, we will convert the CFP into FP using a function defined in the methodology section. FP for each BE is given in Table 21.

\begin{figure}[H]
    \centering
    \includegraphics[width=1\linewidth]{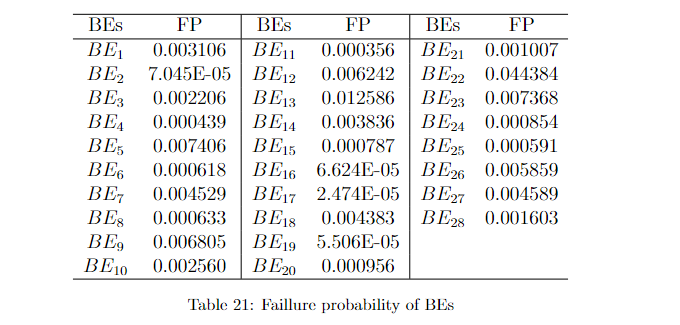}
\end{figure}

    \item Calculating top event probability \\
Using the probabilities associated with each BE and the fault tree structure, we will calculate the probability of the TE. Here the probability of TE is $5.4E-02$.
    
    \item Importance analysis \\
We have utilized the FVI measure to assess the criticality of BEs. The graph below clearly shows that $BE_{13}$ has the highest FVI value, indicating its supreme criticality compared to other BEs.
\begin{figure}[H]
    \centering
    \includegraphics[width=0.8\linewidth]{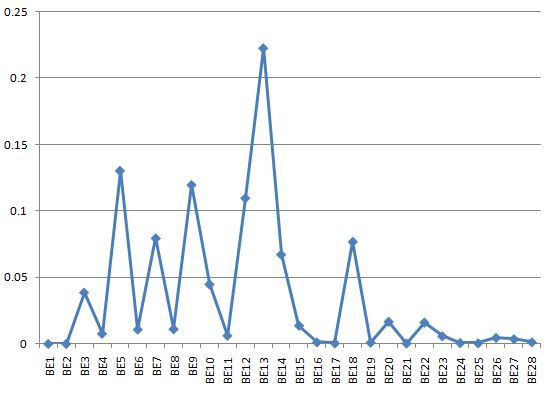}
    \caption{FVI measure of each BE}
\end{figure}   
\end{enumerate}

\subsection{Results and discussion}
The probability of the TE, as calculated by S. Gurgen et al.\cite{GURGEN2023114419}, is 4.86E-02, while our proposed approach produces a slightly higher value $5.4E-02$. The ranking of events based on both methods is provided in Table 22.
\begin{figure}[H]
    \centering
    \includegraphics[width=1\linewidth]{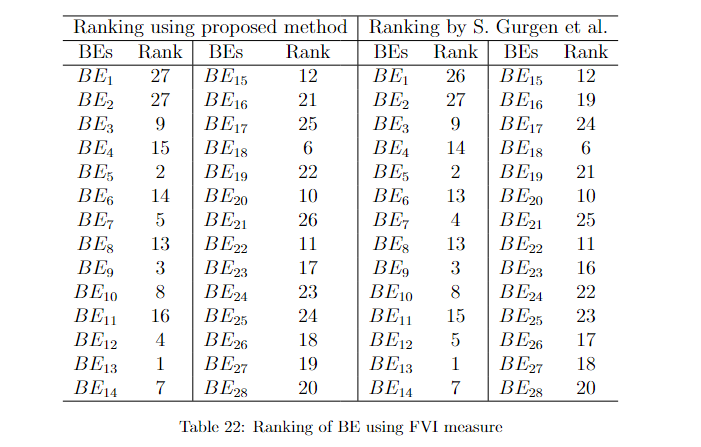}
\end{figure}

\section{Conclusion}
FTA is a powerful tool for risk analysis, offering a clear and visual approach to identifying the underlying causes of an unwanted event and quantifying their likelihood. Its simplicity and ease of use make it a popular choice. Sometimes, there is a lack of information or insufficient data for probability calculations of BEs. Therefore, using qualitative data is a better choice. We are utilizing a fuzzy-based approach to estimate the probability of TE. In this study, we employed the IVFFTA approach to calculate the probability of TE. To do this, we collected expert opinions in linguistic terms and aggregated them using the SAM. We employed the BWM to calculate the weights of the criteria and subsequently used the weighting scores of the experts to determine their weights.

Case study 1, focusing on chemical cargo contamination, identified $BE_{53}$ as the most critical event, with $BE_6$ also posing a significant risk. Furthermore, we calculated the probability of the TE, which is $2.834E-04$. In case study 2, we investigated the root causes of the event loss of ship steering ability. We then ranked the contributing factors using the FVI measure, our analysis showed that event $BE_{13}$ is the most critical contributor to this situation. The probability of TE calculated by the proposed approach is $5.4E-02.$

\end{document}